\documentstyle[12pt,fullpage,epsfig]{article}

\newcommand{\marke}[1]{\protect\label{#1} \fbox {\tt #1} \quad }
\renewcommand{\marke}{\label}

\newcommand{\eq}[1]{Eq.\,(\ref{eq#1})}
\newcommand{\fig}[1]{Fig.\,\ref{fig#1}}

\newcommand{\bE}{{\rm\bf E}}
\newcommand{\bD}{{\rm\bf D}}

\title{Inverse Quantum Zeno Effect in Quantum Oscillations}

\author{Alexander D.\ Panov\\
{\small Skobeltsyn Institute of Nuclear Physics,}\\
{\small Moscow State University, Moscow 119899, Russia}
\thanks{E-mail address: {\tt a.panov@relcom.ru}}
}
\date{}

\begin{document}


\maketitle

\sloppy

\begin{abstract}
It is shown that inverse quantum Zeno effect (IZE) may exist in a
three-level system with Rabi oscillations between discrete atomic
states.  The experiment to observe IZE in such a system is proposed.
\end{abstract}

\begin{center}
\end{center}

\section{Introduction}
\marke{INTRO}

It is known that frequently repeated discrete quantum measurements
can hinder quantum transitions. This phenomenon is known as quantum
Zeno effect (QZE) \cite{ZEN_SUDARSHAN77A,ZEN_SUDARSHAN77B}. This
effect was found experimentally in systems with forced Rabi
oscillations between discrete atomic levels \cite{ZEN_ITANO90} and in
spontaneously decaying systems \cite{ZEN_FISHER01A}. It was shown
also that there are regimes when repeated discrete measurements can
accelerate spontaneous decay
\cite{ZEN_BERNARDINI93,ZEN_PASCAZIO98,ZEN_PASCAZIO01A}, and this
phenomenon was found experimentally as well \cite{ZEN_FISHER01A}.
This effect is known as anti Zeno effect or inverse Zeno effect
(IZE).

In the present paper the definitions for QZE and IZE are admitted in 
agreement with that introduced by P.~Facchi and S.~Pascazio 
\cite{ZEN_PASCAZIO01C} with one modification. Let the initial pure 
state of a system with Hamiltonian $H$ be $\rho_0$ and the survival 
probability be $P(t) = {\rm Tr}[\rho_0 \rho(t)]$. Consider the 
evolution of system under the effect of an additional interaction, so 
that the total Hamiltonian reads
$$
   H_K = H + H_{\rm meas}(K),
$$
where $K$ is a set of parameters and $H_{\rm meas}(K=0) = 0$.  $H$ is
a full Hamiltonian of the system containing interaction terms, and
$H_{\rm meas}(K)$ should be considered as an additional Hamiltonian
performing the measurement. The term $H_{\rm meas}(K)$ may describe a
chain of ideal discrete quantum measurements that are represented by
reductions (collapses) of state of the system as a special case of
interaction.  The system displays QZE if there exist an interval
$I^{(K)} = [t_1^{(K)},t_2^{(K)}]$ such that
\begin{equation}
   P^{(K)}(t) > P(t),\quad \forall t \in I^{(K)},
   \marke{eq2}
\end{equation}
and the system displays IZE if there exist an interval $I^{(K)}$ such
that
\begin{equation}
   P^{(K)}(t) < P(t),\quad \forall t \in I^{(K)}.
   \marke{eq3}
\end{equation}
Here $P^{(K)}(t)$ and $P(t)$ are the survival probabilities under the
action of $H_K$ and $H$, respectively, and it is required
\begin{equation}
   t^{(K)}_2 \leq T_P,
   \marke{eq4}
\end{equation}
where $T_P$ is the Poincar\'e time. The modification of this
definitions is the following. In the addition to the definitions
Eqs.~(\ref{eq2},\ref{eq3},\ref{eq4}) it is required:
\begin{list}{}{}
   \item[(i)]
   The Hamiltonian of measurement $H_{\rm meas}(K)$ is to be
   time-independent or periodical with the period less then the
   Poincar\'e time of the system.
\end{list}
The meaning of condition (i) is the following. Johann von Neumann
proved the following proposition \cite[Chapter V.2]{NEUMANN32}. Using
a sequence of frequently repeated measurements represented by
time-dependent projections it is possible to force the quantum system
pass through arbitrary definite sequence of states.  Particularly, it
is possible to satisfy the conditions of Eqs.~(\ref{eq3},\ref{eq4})
which define IZE. But such a situation does not appear to be IZE,
rather it is a dynamical version of usual quantum Zeno effect
(dynamical quantum Zeno effect, DQZE).  Though DQZE is considered as
``anti-Zeno paradox'' some times \cite{ZEN_BALACHANDRAN01}, such
interpretation seems to be misleading. The condition (i) is intended
to avoid such misunderstands.

It was pointed out many times (see \cite{ZEN_PASCAZIO01A} and
references herein) that both QZE and IZE can be obtained for a
genuinely unstable system, whose Poincar\'e time is infinite.  On the
other hand, the possibility of IZE for an oscillating quantum
mechanical system, whose Poincar\'e time is finite, was not reported
up to now.  Actually, only QZE is possible in two-level quantum
mechanical oscillating systems.  But it is not generally valid in
multilevel systems with the number of levels more than two. In the
present paper a three-level oscillating system with finite Poincar\'e
time exhibiting IZE is constructed.

\section{Interaction picture for evolution interrupted by
measurements}

\marke{REDUCT}

Before discussing of the main subject we introduce the interaction
picture formalisms for the problem of quantum evolution interrupted by
discrete measurements.  Let $H = H_0 + V$ be a Hamiltonian of a
system $S$ and $\rho(t)$ be the density operator of this system.  Let
$\{P_i\}$, $P_i^2 = P_i$, $\sum_i P_i = 1$ be a complete set of
projection operators. This set of projectors represents an
instantaneous reduction of system state following an ideal quantum
measurement.  The change of state during the measurement is
\begin{equation}
   \rho' = \sum_i P_i \rho P_i \equiv \hat R \rho,
   \marke{eq5}
\end{equation}
where $\rho$ is the state before the measurement and $\rho'$ is the
state after the measurement. Let $D(t)\,dt$ be the probability of
carrying out the measurement on the system during time interval
$(t,t+dt)$.  Then, it is easy to prove that the state of system is
governed by the Lindblad equation
\begin{equation}
   \frac{d\rho}{dt} =
   - \frac{i}{\hbar} [H_0 + V, \rho]
   -\frac{1}{2} D(t) \sum_i[P_i,[P_i,\rho]].
   \marke{eq6}
\end{equation}
Let $\rho_I(t)$ and $V_I(t)$ be the state and the Hamiltonian of the
system in interaction picture:
\begin{eqnarray}
   \rho_I(t) &=& \exp\left(\frac{i}{\hbar} H_0 t\right) \rho(t)
                 \exp\left(-\frac{i}{\hbar} H_0 t\right)
   \marke{eq7}\\
   V_I(t) &=& \exp\left(\frac{i}{\hbar} H_0 t\right) V
                 \exp\left(-\frac{i}{\hbar} H_0 t\right).
   \marke{eq8}
\end{eqnarray}
Further, let
\begin{equation}
   [P_i, H_0] = 0\quad \forall i.
   \marke{eq9}
\end{equation}
By direct substitution of Eqs.~(\ref{eq7},\ref{eq8}) in \eq{6} and
accounting for \eq{9} it is not hard to prove that
\begin{equation}
   \frac{d\rho_I}{dt} =
   - \frac{i}{\hbar} [V_I, \rho_I]
   -\frac{1}{2} D(t) \sum_i[P_i,[P_i,\rho_I]].
   \marke{eq10}
\end{equation}
Equation (\ref{eq10}) is a generalization of Lindblad equation (\ref{eq6})
for the interaction picture of evolution.

Let consider the evolution of the system during the time interval
$(0, t)$.  Let $t_0, t_1, \dots, t_n$ be moments of time such that
$t_0=0 < t_1 < \dots <t_{n-1} < t_n=t$. Then \eq{6} and \eq{10} are
also valid for the singular distribution $D(t)$:
$$
   D(t) = \sum_{k=0}^n \delta(t - t_k).
$$
where $\delta()$ is the Dirac's delta-function.  This special
distribution $D(t)$ describes the sequence of measurements at the
definite moments of time $t_0, t_1, \dots, t_n$. It is not hard to
understand that the solution of \eq{10} for this special $D(t)$ may
be written as
\begin{equation}
   \rho_I(t) = \hat R \hat U_I(t_n,t_{n-1}) \cdots
               \hat R \hat U_I(t_1,t_0)
               \hat R \rho(t_0),
   \marke{eq12}
\end{equation}
where the superoperator of reduction $\hat R$ is defined by \eq{5}
and the superoperator of evolution
\begin{equation}
   \hat U_I(t'', t')\rho = U_I(t'',t') \rho U_I^+(t'',t')
   \marke{eq12p}
\end{equation}
is defined by the solution of Shr\"odinger equation in the
interaction picture without measurements:
$$
   \frac{d\rho_I}{dt} =
   - \frac{i}{\hbar} [V_I, \rho].
$$
If the system was prepared in the pure eigenstate $|\Psi_0\rangle$ of
Hamiltonian $H_0$ at the initial moment of time $t_0=0$, then it is
easily shown that the survival probability $P(t)$ would be
\begin{equation}
   P(t) = \langle \Psi_0 | \rho_I(t) | \Psi_0 \rangle,
   \marke{eq14}
\end{equation}
where $\rho_I(t)$ is defined by \eq{12}.

\section{Model system}
\marke{MODEL}

Let consider three-level atom with free Hamiltonian $H_0$ and
eigenstates $|0\rangle,|1\rangle,|2\rangle$:
\begin{eqnarray}
   H_0|j\rangle &=& \hbar\omega_j|j\rangle, \quad j = 0,1,2
   \marke{eq15}\\
   \omega_{ij} &=& \omega_i - \omega_j, \quad i \ne j.
   \nonumber
\end{eqnarray}
Let the initial state of the atom be
$
   |\Psi_0\rangle = |0\rangle.
$
The atom interacts with the classical electric field consisting of
two components being in resonance with the transitions $\omega_{10}$
and $\omega_{21}$, respectively:
\begin{equation}
   \bE(t) =
   \bE_{10} e^{i\omega_{10}t} + \bE_{10}^* e^{-i\omega_{10}t} +
   \bE_{21} e^{i\omega_{21}t} + \bE_{21}^* e^{-i\omega_{21}t},
   \marke{eq18}
\end{equation}
where $\bE_{10}$ and $\bE_{21}$ are arbitrary complex amplitudes of
fields. The interaction of electric field with the atom is
\begin{equation}
   V = -\bD \bE,
   \marke{eq20}
\end{equation}
where $\bD$ is the operator of dipole moment of the atom. The
following relations are admitted to be valid:
\begin{eqnarray}
   |\omega_{ij}| &\gg& |V_{mn}| \quad \forall i,j,m,n;
   \marke{eq21}\\
   |\omega_{ij} - \omega_{kl}| &\gg& |V_{mn}| \quad
   \forall i,j,k,l : \{i,j\} \ne \{k,l\}, \quad \forall m,n,
   \marke{eq22}
\end{eqnarray}
where $V_{mn} = \langle m | V | n \rangle$.

Let consider the evolution of atom during time interval $(0,t)$.
Suppose that the measurement $\hat R$ is carried out on the atom at
the moments $t_0,t_1,\dots t_n$ where $t_k = k\Delta t,\Delta t =
t/n$, $n = 1, 2, 3, \dots$ . The measurement $\hat R$ is intended to
determine whether the atom is on the level $|2\rangle$ or not.
Therefore, the superoperator $\hat R$ reads
\begin{equation}
   \hat R \rho = P_{01} \rho P_{01} + P_2 \rho P_2,
   \marke{eq23}
\end{equation}
where
\begin{equation}
   P_{01} = {\rm diag}(1,1,0),\quad
   P_2    = {\rm diag}(0,0,1).
   \marke{eq24}
\end{equation}

Let find the probability that the atom is at the state $|0\rangle$ at
time $t$. This probability is the survival probability $P(t)$ \eq{14}
for $|\Psi_0\rangle = |0\rangle$. To determine $\rho_I(t)$ one can
use \eq{12}.  Further, $\hat R$ in \eq{12} is already known from
\eq{23}.  Consequently, $\hat U_I(t'',t')$ is to be calculated.

Let $a(t)$ be a three-dimensional complex vector, $a = [a_0, a_1,
a_2]$.  Consider the equation
\begin{equation}
   \frac{da(t)}{dt} = -\frac{i}{\hbar}V_I(t) a(t)
   \marke{eq25}
\end{equation}
with the initial conditions defined for the time $t'$: $a(t') =
[a^0_0,a^0_1,a^0_2]$.  In \eq{25} $V_I(t)$ is the interaction picture
Hamiltonian for the free Hamiltonian \eq{15} and the interaction
\eq{20}.  The solution of \eq{25} for the time $t''$ may be written as
\begin{equation}
   a(t'') = U_I(t'',t')a(t'),
   \marke{eq26}
\end{equation}
where $U_I(t'',t')$ is the evolution operator that is needed for
calculation of $\hat U_I(t'',t')$ by \eq{12p}.  Using
Eqs.~(\ref{eq18},\ref{eq20}) and the rotating wave approximation
(which is right under the conditions (\ref{eq21},\ref{eq22})), one
can rewrite \eq{25} as
\begin{equation}
   \frac{d}{dt}
   \left(
   \begin{array}{ccc}
      a_0\\ a_1\\ a_2
   \end{array}
   \right)
   =  -i
   \left(
   \begin{array}{ccc}
      0 & \Omega_{01}e^{i\varphi_{01}} & 0 \\
      \Omega_{01}e^{-i\varphi_{01}}
      & 0 &
      \Omega_{12}e^{i\varphi_{12}}\\
      0 & \Omega_{12}e^{-i\varphi_{12}} & 0 \\
   \end{array}
   \right)
   \left(
   \begin{array}{ccc}
      a_0\\
      a_1\\
      a_2
   \end{array}
   \right),
   \marke{eq27}
\end{equation}
where the notations
$$
   \Omega_{01}e^{i\varphi_{01}}
   = -\langle 0|\bD|1 \rangle \bE_1 / \hbar,
   \quad
   \Omega_{12}e^{i\varphi_{12}}
   = -\langle 1|\bD|2 \rangle \bE_2 / \hbar,
$$
were introduced. The values $\Omega_{01}$ and $\Omega_{12}$ are
considered to be positive real numbers. The evolution operator
$U_I(t'',t')$ may be obtained by solution \eq{27} and comparison the
results with \eq{26}:
\begin{equation}
   \begin{array}{l}
      \displaystyle
      U_I(t'',t') = \\
      \ \\
      \left(
      \begin{array}{ccc}
         \displaystyle
         \frac{\Omega_{12}^2 + \Omega_{01}^2\cos\alpha}{\Omega^2}
         &
         \displaystyle
         -i\frac{\Omega_{01}}{\Omega}e^{i\varphi_{01}}\sin\alpha
         &
         \displaystyle
         -\frac{\Omega_{01}\Omega_{12}}{\Omega^2}
         e^{i(\varphi_{01} + \varphi_{12})} (1 - \cos\alpha)
         \\
         \displaystyle
         -i\frac{\Omega_{01}}{\Omega} e^{-i\varphi_{01}} \sin\alpha
         &
         \displaystyle
         \cos\alpha
         &
         \displaystyle
         -i\frac{\Omega_{12}}{\Omega} e^{i\varphi_{12}} \sin\alpha
         \\
         \displaystyle
         -\frac{\Omega_{01}\Omega_{12}}{\Omega^2}
         e^{-i(\varphi_{01} + \varphi_{12})} (1 - \cos\alpha)
         &
         \displaystyle
         -i\frac{\Omega_{12}}{\Omega}e^{-i\varphi_{12}}\sin\alpha
         &
         \displaystyle
         \frac{\Omega_{01}^2 + \Omega_{12}^2\cos\alpha}{\Omega^2}
      \end{array}
      \right),
   \end{array}
   \marke{eq29}
\end{equation}
where
$$
   \Omega = \sqrt{\Omega_{01}^2 + \Omega_{02}^2}\, ; \quad
   \alpha = \Omega(t'' - t').
$$
Now both values, $\hat R$ (from Eqs.~(\ref{eq23},\ref{eq24})) and
$\hat U_I(t'',t')$ (from Eqs.~(\ref{eq12p},\ref{eq29})), are known,
and the survival probability for state $|0\rangle$ may be
calculated by Eqs.~(\ref{eq12},\ref{eq14}).

\section{Results of calculations and discussion}
\marke{DISCUSS}

Firstly, let us discuss the mechanisms of IZE in this system 
qualitatively.  For the beginning suppose that the measurements 
are absent at all.  Since the initial state of system $|\Psi_0\rangle 
= |0\rangle$ was pure at the moment $t_0 = 0$ so it will be pure in 
future, and the evolution will be governed only by the operator 
$U_I(t'',t')$:
$
   |\Psi_I(t)\rangle = U_I(t,0)|0\rangle.
$
Suppose $\Omega_{12} = 0$. It is seen from \eq{29} that the evolution 
of the atom is reduced to the usual Rabi 
oscillations\footnote{Hereafter we suppose $\varphi_{01} = 
\varphi_{12} = 0$ since these phases do not effect the probabilities 
to find any atomic state at any time $t > 0$ if the initial state of 
atom was $|0\rangle$.}:
$$
   a_0(t) =  \cos\Omega_{01} t;\quad
   a_1(t) = -i\sin \Omega_{01} t;\quad
   a_2(t) = 0.
$$
In the converse case, $\Omega_{12} \gg \Omega_{01}$, \eq{29} shows 
that the initial state is "frozen":
$$
   \forall t:\quad
   a_0(t) \approx 1;\quad
   a_1(t) \approx 0;\quad
   a_2(t) \approx 0.
$$
Generally speaking, the transition between states $|1\rangle$ and 
$|2\rangle$ hinders the transition between states $|0\rangle$ and 
$|1\rangle$. This is well-known phenomenon 
\cite{ZEN_PERES80A,ZEN_KRAUS81,ZEN_PASCAZIO01C} which is considered 
as a Zeno-like effect. However, if the transition $|1\rangle \to 
|2\rangle$ itself is continuously observed by frequent measurements 
represented by \eq{23}, then the transition $|1\rangle \to |2\rangle$ 
will be "frozen" by usual QZE, and the mentioned above Zeno-like 
effect will be hindered by this usual Zeno effect. Rabi transition 
$|0\rangle \to |1\rangle$ becomes possible followed by the state 
$|0\rangle$ "defreezes". And this is IZE.

To represent the detailed calculations, $\Omega_{12} = 
\Omega_{01}\sqrt{15}$ is chosen. Hence $\Omega = 4\Omega_{01}$ and it 
is seen from \eq{29} that the Poincar\'e time of the system is
\begin{equation}
   T_P = 2\pi / \Omega = \pi / (2\Omega_{01}).
   \marke{eq34}
\end{equation}
The solid line on \fig{1} represents the ``free'' evolution of 
the atom during one Poincar\'e time:  both resonant components of 
electric field is switched on, but measurements is switched off. 
Dashed lines represent the evolution with different numbers of 
measurements during the interval $t \in (0,T_P)$; the number near the 
line is the number $n$ as in \eq{12}.  It is seen from \fig{1} that 
IZE takes place in the exact accordance with the definition of IZE by 
Eqs.~({\ref{eq3},\ref{eq4}) and condition (i).  Moreover, the 
evolution of the atom tends to the free Rabi transition between 
levels $|0\rangle$ and $|1\rangle$ with frequency $2\Omega_{01}$ as 
$n\to\infty$, as one should expect.

The model three-level system with double Rabi transition and 
measurements described in the present paper may be realized in the 
experiment similar to Itano and collaborators QZE-experiment with 
simple Rabi transition \cite{ZEN_ITANO90}.  The levels 
$|0\rangle,|1\rangle,|2\rangle$ of the atom may correspond to fine or 
hyperfine structure. Rabi transitions between these levels may be 
forced by ultra high frequency or radio frequency fields.  In the 
addition the forth level $|3\rangle$ should be involved such that it 
should be higher than level $|2\rangle$ and the transition 
$|3\rangle\to|2\rangle$ should be a no-forbidden optical transition.  
The state $|3\rangle$ is to decay onto the state $|2\rangle$ much 
faster than all Rabi transitions involved to the experiment. A short 
$\pi$-pulse of laser tuned to the resonance with the transition 
$|3\rangle\to|2\rangle$ will simulate the measurement $\hat R$, 
\eq{23}. The probability of finding the atom in the state $|0\rangle$ 
at the end of evolution may be measured by the usual way 
\cite{ZEN_ITANO90}.


The author acknowledges the fruitful discussions with P.~Facchi, 
M.~B.~Mensky, and S~.Pascazio and is grateful to V.~A.~Arefjev for 
the help in preparation of the paper.



\begin{figure}[h]
   \begin{center}
      \epsfig{file=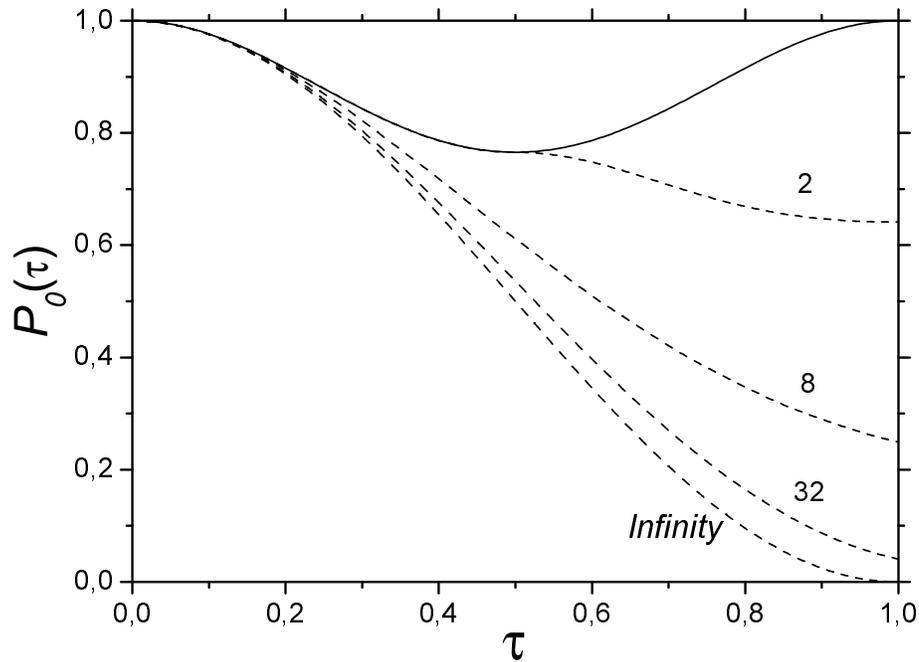,width=15cm}
   \end{center}
   \caption{Inverse Zeno effect in a three-level atom with double
   Rabi transition. $\tau = t/T_P$, where Poincar\'e time $T_P$ is
   defined by \protect\eq{34}; $P_0(\tau)$ is the probability to find
   the atom in the initial state $|0\rangle$. Solid line reperesents
   the evolution of atom without measurements, dashed lines represent
   the evolution of atom with different number of measurements during
   the interval $\tau \in (0,1)$.}
   \label{fig1}
\end{figure}

\end{document}